**Title:** Improving CCTA based lesions' hemodynamic significance assessment by accounting for partial volume modeling in automatic coronary lumen segmentation

**Authors:**

Moti Freiman* PHD, Hannes Nickisch PHD, Sven Prevrhal PHD, Holger Schmitt PHD, Mani Vembar, Pál Maurovich-Horvat MD PHD MPH,  Patrick Donnelly MD MB BCh BAO FRCP, and Liran Goshen PHD

*Corresponding Author (Address correspondence)

Moti Freiman, Philips Medical Systems Technologies Ltd.

Advanced Technologies Center, Building No. 34

P.O. Box 325

Haifa 3100202

Israel

E-mail: moti.freiman@philips.com

**Funding information:** None.

**Article type:** Original research

Accepted for publication at the journal "Medical Physics", Jan, 2017.

**Abstract:**

**_Purpose:_**  The goal of this study was to assess the potential added benefit of accounting for partial volume effects (PVE) in an automatic coronary lumen segmentation algorithm that is used to determine the hemodynamic significance of a coronary artery stenosis from coronary computed tomography angiography (CCTA).

**_Materials and methods:_**

Two sets of data were used in our work: (1) multi-vendor CCTA datasets of 18 subjects from the MICCAI 2012 challenge with automatically generated centerlines and 3 reference segmentations of 78 coronary segments and (2) additional CCTA datasets of 97 subjects with132 coronary lesions, that had invasive reference standard FFR measurements.  We extracted the coronary artery centerlines for the 97 datasets by an automated software program followed by manual correction if required. An automatic machine-learning based algorithm segmented the coronary tree with and without accounting for the PVE.  We obtained CCTA based FFR measurements using a flow simulation in the coronary trees that were generated by the automatic algorithm with and without accounting for PVE. We assessed the potential added value of PVE integration as a part of the automatic coronary lumen segmentation algorithm by means of segmentation accuracy using the MICCAI 2012 challenge framework and by means of flow simulation overall accuracy, sensitivity, specificity, negative and positive predictive values and the receiver operated characteristic (ROC) area under the curve. We also evaluated the potential benefit of accounting for PVE in automatic-segmentation for flow-simulation for lesions that were diagnosed as obstructive based on CCTA which could have indicated a need for an invasive exam and revascularization.

***Results:***

Our segmentation algorithm improves the maximal surface distance error by ~39% compared to previously published method on the 18 datasets from the MICCAI 2012 challenge with comparable Dice and mean surface distance. Results with and without accounting for PVE were comparable. In contrast, integrating PVE analysis into an automatic coronary lumen segmentation algorithm improved the flow simulation specificity from 0.6 to 0.68 with the same sensitivity of 0.83. Also, accounting for PVE improved the area under the ROC curve for detecting hemodynamically significant CAD from 0.76 to 0.8 compared to automatic segmentation without PVE analysis with invasive FFR threshold of 0.8 as the reference standard. Accounting for PVE in flow simulation to support the detection of hemodynamic significant disease in CCTA-based obstructive lesions improved specificity from 0.51 to 0.73 with same sensitivity of 0.83 and the area under the curve from 0.69 to 0.79. The improvement in the AUC was statistically significant (N=76, Delong's test, p=0.012).

***Conclusion:*** Accounting for the partial volume effects in automatic coronary lumen segmentation algorithms has the potential to improve the accuracy of CCTA-based hemodynamic assessment of coronary artery lesions.

## 1. <u>Introduction:</u>

Coronary artery disease (CAD) is the single leading cause of death worldwide, accounting for 11.2% of all deaths globally in 2011.[1] Among the non-invasive tests available for patients with suspected CAD, Coronary Computed Tomography Angiography (CCTA) is a rapidly evolving technique to rule out CAD due to its high negative predictive value.[2] However, compared to other non-invasive functional tests available, CCTA provides mainly an anatomical characterization of the coronary lesions rather than an assessment of their hemodynamic significance.[3] Recent studies suggest that the hemodynamic significance of a CT coronary stenosis by means of Fractional Flow Reserve (FFR, i.e. the ratio between the pressure after a lesion and the normal pressure) can be assessed from CCTA data using flow simulations. Early reports have demonstrated that this strategy can improve the specificity of CCTA for the detection of CAD. [4–8]

Non-invasive assessment of the hemodynamic significance of a coronary stenosis from CCTA requires a three-dimensional coronary tree model to perform flow simulation calculations. Such models are commonly generated by time-consuming manual refinement of automatic coronary lumen segmentation algorithm results. For example, Coenen et al.[8] report that the time required for semi-automatic coronary segmentation for flow simulation varies depending on the extent of atherosclerotic disease, with a range of 30-120 minutes per patient. This time-consuming semi-automatic segmentation step may therefore impede the routine clinical utilization of flow simulation as part of the CCTA exam.

The challenge of automatic coronary segmentation from CCTA in particular and vessel segmentation in general, were addressed by many researchers in the past few years.[9–11] The publicly available MICCAI 2012 coronary segmentation challenge database[12] allows the comparison of multiple coronary lumen segmentation algorithms on the same basis.

Graph-based algorithms that incorporate some anatomical prior knowledge show promising results in segmentation of tubular structures. For example, Kang et al[13] show how to obtain a globally optimal surface of tube-like structures with validation on phantom CT images and Gopalkrishna et al[14] present an algorithm to segment the atrium wall by using globally optimal graph-based optimization. Specifically, Lugauer et al[15,16] obtained the best reported results on the MICCAI 2012 coronary segmentation challenge database[12] by combining a machine-learning based boundary detection, with a graph min-cut based optimal surface generation.

The MICCAI 2012 evaluation methodology was focused, however, on the anatomical agreement between the automatic and expert manual segmentations. [11] With current interest in hemodynamic significance assessment from CCTA, it is important to assess the impact of automatic segmentation performance on the CCTA-based hemodynamic significance assessment.

The accuracy of automatic coronary segmentation algorithms is dependent on the image quality of the final dataset and on the contrast attenuation between the lumen and the neighboring region which may include calcified and non-calcified plaque. In addition, overall image resolution can affect the accuracy of the automatic segmentation results. Specifically, the finite resolution of imaging scanners and blurring involved in the reconstruction which are integrated into the overall system Point Spread Function (PSF) may lead to an overestimation of lumen area in vessels with small lumen diameter which is known as the partial-volume effect (PVE).[17,18] The result of the PVE-related overestimation of the lumen area may cause underestimation of the lesion's hemodynamic significance. Fig. 1 illustrates the effect of PVE on estimating vessel radius using full-width half maximum rule[19] on 2D vessel profiles with varying stenosis percentage due to the presence of non-calcified plaque.

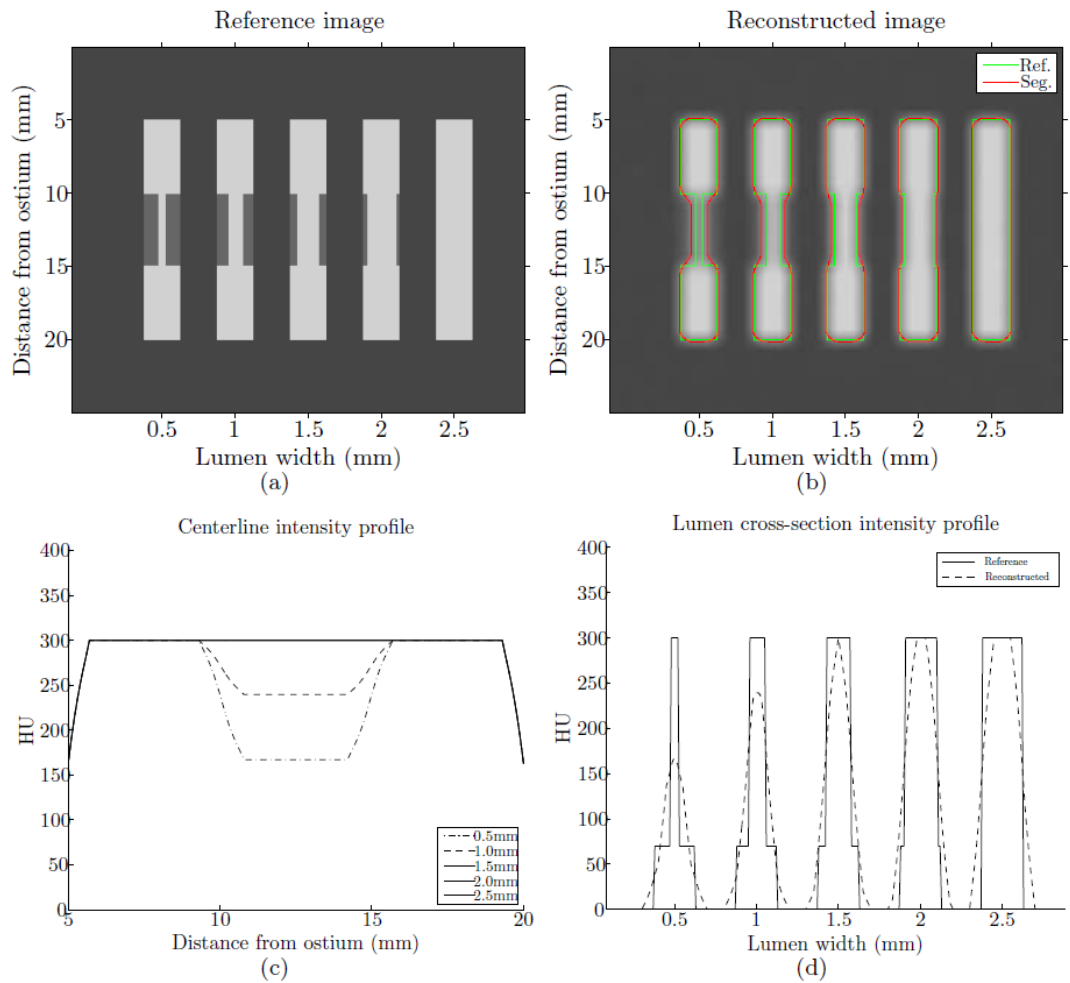

**Fig. 1:** The effect of PVE on estimating vessel radius using the full-width half maximum rule[19] on 2D vessel profiles with varying stenosis percentage due to the presence of non-calcified plaque. (a)The reference image, including ideal vessel profiles with varying stenosis percentage due to the presence of soft-plaque. (b) The reconstructed image, with the reference segmentation in green and full-width half maximum (FWHM) segmentation in red). Note the over-estimation of the lumen diameter due to the partial volume effect. (c) The vessel intensity profile. (d) The vessels cross-sectional intensity profile. The HU reduction in the centerline intensity profile can be used to determine locations that affected by partial volume effect.

Recently, a new automatic coronary lumen segmentation algorithm was presented at the SPIE 2016 Medical Imaging meeting by Freiman et al that accounts specifically for potential PVE in order to improve the performance of coronary lesions' hemodynamic significance assessment from CCTA data. [20]

The goal of this work was to assess the potential added value of accounting for the PVE by an automatic coronary segmentation algorithm in the assessment of the hemodynamic significance of a CT coronary stenosis by flow simulation.

## 2.  Materials and methods:

### 2.A. Datasets:

We used two data sets as follows.

The first dataset was the publicly available MICCAI 2012 coronary artery segmentation challenge database. The database consists of 48 CCTA datasets that were acquired from a representative selection of CAD symptomatic patients using several cardiac CT scanners from different vendors with varying protocols and reconstruction algorithms. Reference cross-sectional contours representing the lumen segmentation were annotated by three different experts. The first 18 are available for algorithm training while the remaining 30 datasets were for testing only. In addition, coronary centerlines were provided to initialize the segmentation. In all of our experiments with the MICCAI 2012 database we used the centerlines provided by the method of Goldenberg et al. [21] We refer the reader to [11,12] for a detailed description of the data and evaluation methodology.

Since full evaluation using this framework require software for stenosis detection and quantification which is beyond the scope of this contribution, we limited our evaluation using the MICCAI 2012 data to the training dataset for which reference lumen contours were available.

The second dataset consists of CCTA data of 132 coronary lesions that were retrospectively collected from the medical records of 97 subjects who underwent a CCTA and invasive coronary angiography with invasive FFR measurements due to suspected CAD. CCTA data was acquired using either a Philips Brilliance iCT (gantry rotation time of 0.27 sec.) or Philips Brilliance 64 (gantry rotation time of 0.42 sec.). Acquisition mode was either helical retrospective ECG gating (N=54) or prospectively ECG triggered axial scan (N=43). The kVp range was 80 – 140 kVp and the tube output range was 600 – 1000 mAs for the helical retrospective scans and 200 – 300 mAs for the prospectively ECG trigged scans.

Cross-sectional area (CSA) based stenosis quantification was performed by an expert reader on 132 lesions, out of which 56 were diagnosed as non-obstructive lesions (CSA stenosis less than 50%) and 76 diagnosed as obstructive lesions (CSA stenosis 50%-90%). According to the invasive FFR measurements, 48 lesions were hemodynamically significant (FFR≤0.8) and 84 lesions were non-significant (FFR>0.8).

**2.B. Coronary Lumen Segmentation Algorithm:**

The proposed coronary lumen segmentation algorithm requires the following inputs:

1) The CCTA volume

2) The coronary-artery centerlines

3) The segmentation of the aortic root

The coronary artery centerlines and the aorta segmentation were computed automatically and adjusted manually by a cardiac CT expert (M.V) to account for algorithm inaccuracies using a commercially available software dedicated for cardiac image analysis (Comprehensive Cardiac Analysis, IntelliSpace Portal 6.0, Philips Healthcare).

The coronary lumen segmentation algorithm starts with the analysis of the intensity profile along the coronary centerline to detect regions with small lumen diameter that may be overestimated due to the PVE, followed by estimation of underlying lumen radius, which is then used within a machine-learning based graph-cut algorithm yielding the final segmentation. Fig. 2 presents a schematic flowchart of the proposed algorithm. We describe each step in detail in the following.

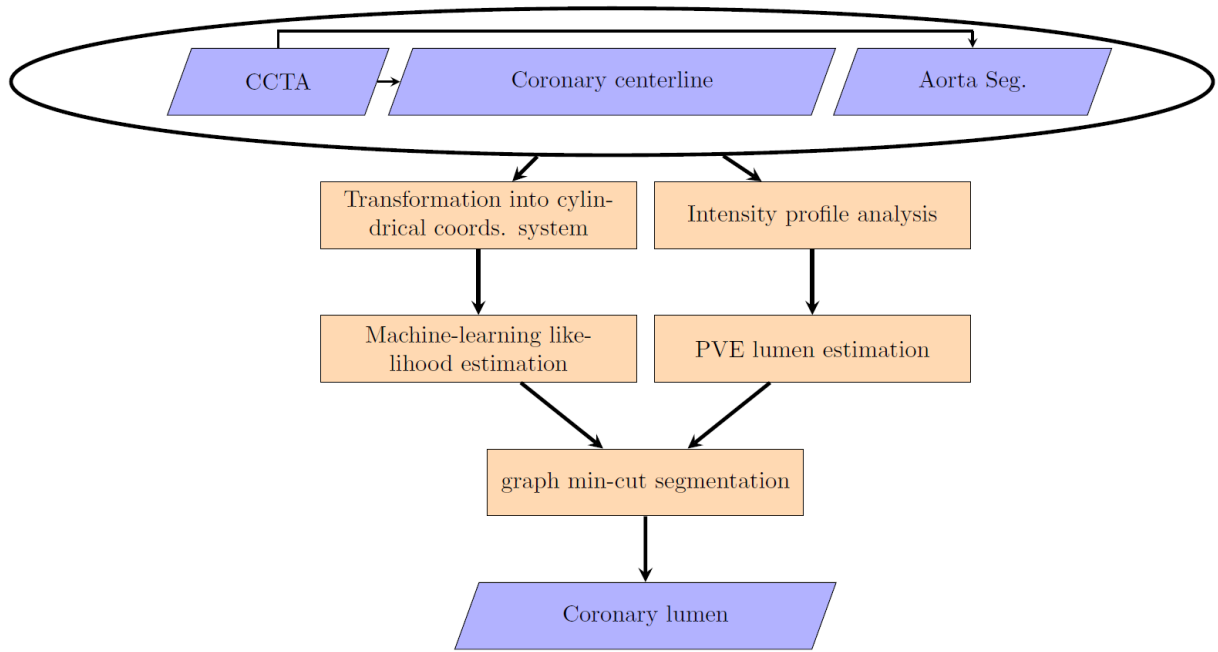

**Fig. 2:** The coronary lumen segmentation algorithm schematic flowchart. The algorithm required the following inputs: 1) the CCTA volume, 2) the coronary artery centerlines, and 3) the aortic root segmentation. The algorithm consists of the following steps: 1) Analysis of the intensity profile along the coronary centerline to detect regions with small diameter lumen that may be overestimated due to the PVE, 2) Estimation of underlying lumen radius, 3) Transformation into a cylindrical coordinate system around the coronary centerline, 4) machine-learning based likelihood estimation, and; 5) final segmentation by the graph-cut segmentation framework.

2.B.1. Partial volume effect artifacts detection and estimation:

The goal of this component of the algorithm is to determine locations along the coronary centerline that might be subject to the PVE for a given a cardiac CT angiography volume ($I$) with a coronary centerline ($C$). We first model the expected intensity profile along the coronary centerline by a polynomial function $I_p(c)$ parameterized over the distance between the point $c$ and the beginning of the coronary ostium:

$$I_p(c) = \beta_0 + \sum_{n=1}^{2} \beta_n \cdot dist(c)^n \quad [1]$$

where $dist(c)$ is the centerline curve length from its ostium to the centerline point $c$. [22] We fit the model $I_p(c)$ to the intensity profile along the centerline $I(c)$ using a two-phase robust intensity profile model fitting with outlier detection. First, we fit the model to the HU sampled along the centerline by minimizing a least-squares criterion.

$$\widehat{I_p(c)} = arg \min_{I_p(C)} \sum_{c \in C} \big(I_p(c) - I(c)\big)^2 \quad [2]$$

We used the Student's t-test to identify intensity samples along the centerline that are significantly different (i.e., $p < 0.05$) from the estimated model. We define samples as outliers where the intensity values along the centerline that are 2 standard deviations below the model-based expected intensity.

$$1_{PV}(c) = \begin{cases} 0, & I(c) > I_p(c) - 2\sigma_{I_c} \\ 1, & I(c) \leq I_p(c) - 2\sigma_{I_c} \end{cases} \quad [3]$$

where $\sigma_{I_c}$ is the standard deviation over the differences between the intensity profile along the centerline $I(c)$ and the fitted intensity model $I_p(c)$. These outliers are potentially locations along the coronary centerline that may be affected by the PVE. To obtain a robust centerline intensity profile model we exclude these outliers from the centerline, i.e. $C_{clean} = C \backslash C_{outliers}$ and fit the model again using $C_{clean}$ instead of $C$.

We estimated the actual underlying coronary radius in regions with PVE ($1_{PV}(c) = 1$) by modeling the radius of the coronary lumen at centerline location $c$ as a linear function parameterized over the percentage decrease in lumen intensity at location $c$ compared to the model-based expected intensity:

$$r(c) = 0.5 \left( \alpha \left( 1 - \frac{I(c)}{I_p(c)} \right) + \beta \right) \quad [4]$$

where $r(c)$ is in units of [mm]. We calculate the model coefficients $\alpha, \beta$ by fitting the model to a mathematical phantom simulating the HU reduction in ideal vessel profiles with varying diameter. We set the model parameter values as follows: $\alpha = -2.0mm$ and $\beta = 1.4mm$. Note that this function is defined only for regions with PVE (i.e. $1_{PV}(c) = 1$). Fig. 3 presents the mathematical phantom simulation experimental measurement of the percentage Hounsfield unit (HU) reduction as a function of the coronary diameter along with the fitted model (Eq. 4) used in our method.

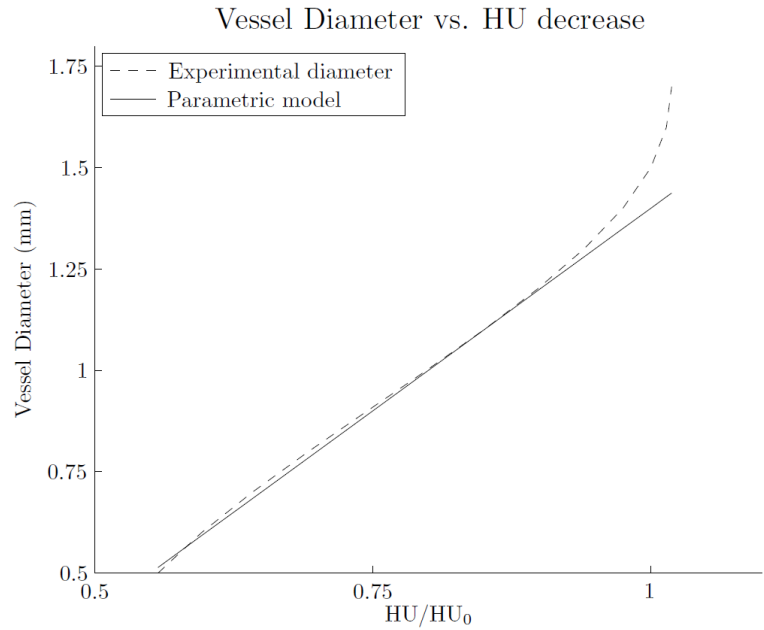

**Fig. 3:** Experimental measurement of the percentage HU reduction as a function of the coronary diameter along with the fitted model (Eq. 4) where HU is the measured HU at the vessel centerline and $HU_0$ is the expected HU at the location without the PVE.

2.B.2. Machine-learning based graph min-cut coronary segmentation:

We formulate the segmentation task as an energy minimization problem over a cylindrical coordinate system[15,16] where the warped volume along the coronary artery centerline is expressed with the coordinate $i$ representing the index of the cross-sectional plane, and $\theta, r$ represent the angle and the radial distance determining a point in the cross-sectional plane:

$$E(X) = \sum_{p \in P} \psi_p(x_p) + \lambda \sum_{p,q \in E} \psi_{p,q}(x_p, x_q) \quad [5]$$

where $P$ is the set of sampled points, $x_p$ is a vertex in the graph representing the point $(i^{x_p}, \theta^{x_p}, r^{x_p})$ sampled from the original CCTA volume, $\psi_p(x_p)$ represents the likelihood of the vertex to belong to the lumen or the background class, $p, q$ are neighboring vertices according to the employed neighboring system $E$, and $\psi_{p,q}(x_p, x_q)$ is a penalty for neighboring vertices belonging to different classes ensure the smoothness of the resulted surface.

We use the K-nearest-neighbor approach[23] to calculate the likelihood of each vertex $x_p$ to belong to the coronary lumen from a large training database consists of rays sampled from cardiac CTA data along with matched binary rays representing the manual segmentation represents the distribution of coronary rays and their segmentations as follows. We first obtain a set of K similar rays by means of $L^2$ norm between the rays' intensity profiles using Muja and Lowe's approximated K-nearest-neighbor algorithm. [24] We then use a kernel density estimator to calculate the probability of $x_p$ belonging to the lumen:

$$Pr_d(x_p \in Lumen) = \frac{\sum_{k=1}^{K} w\left(I((i^{x_p}, \theta^{x_p}, R), I'(i^k, \theta^k, R)\right) \cdot \delta\left(x_p, S(i^k, \theta^k, R)\right)}{\sum_{k=1}^{K} w\left(I((i^{x_p}, \theta^{x_p}, R), I'(i^k, \theta^k, R)\right)} \quad [6]$$

where $R$ is a set of radial distances, $I(i^{x_p}, \theta^{x_p}, R)$ is the sampled ray that include the point $x_p$ in the new volume, $I'\left(i^k, \theta^k, R\right)$ is the ray from the training set, $\delta\left(x_p, S\left(i^k, \theta^k, R\right)\right)$ is an indicator function that indicates whether the point $x_p$ is labeled with 1 on the binary ray $S\left(i^k, \theta^k, R\right)$ corresponding to the $I'\left(i^k, \theta^k, R\right)$ ray in the training data, $K$ is the number of closest rays to be used, and $w\left(I(i^{x_p}, \theta^{x_p}, R), I'\left(i^k, \theta^k, R\right)\right)$ is a weighting function that is used to weight the contribution of each training ray according to its distance from the test ray:

$$w\left(I(i^{x_p}, \theta^{x_p}, R), I'\left(i^k, \theta^k, R\right)\right) = \exp\left(-\lambda \left\| I(i^{x_p}, \theta^{x_p}, R) - I'\left(i^k, \theta^k, R\right) \right\|_2^2\right) \quad [7]$$

To account for the PVE, we adjust the probability of points along the ray calculated with Eq. 6 to reflect the estimated radius, calculated as described in Sec. 2.B.1 for every ray belongs to planes which we identified with potential small lumen diameter according to Eq. 3. First, we define the probability of each point $x_p$ to belong to the lumen according to the estimated radius at the cross section $i^{x_p}$ calculated as described in Sec. 2.B.1:

$$Pr_{PV}(x_p \in lumen) = \begin{cases} 0, & r^{x_p} > r' \\ 1, & r^{x_p} \leq r' \end{cases} \quad [8]$$

where $r^{x_p}$ is the radial distance of the point $x_p$ and $r'$ is the estimated radius at the cross section $i^{x_p}$.

Next, we combine the two probabilities together:

$$Pr(x_p \in lumen) = \begin{cases} Pr_d(x_p \in Lumen), & 1_{PV}\left(c(x_p)\right) = 0 \\ Pr_{PV}(x_p \in lumen), & 1_{PV}\left(c(x_p)\right) = 1 \end{cases} \quad [9]$$

where $c(x_p)$ is the centerline location that the sampling point $x_p$ belongs to. We also adjust vertices represent calcified plaque as determined by HU above a fixed threshold to have a high background probability value.

Finally we assign:

$$\psi_p(x_p) = -\log Pr(x_p \in lumen) \quad [10]$$

We use the L$^2$ intensity difference regularization term to encourage a smooth surface result:

$$\psi_{p,q}(x_p, x_q) = \exp\left(-\frac{(I(x_p)-I(x_q))^2}{\sigma_c(x_p)}\right) \cdot \exp\left(-d(x_p, x_q)^2\right) \quad [11]$$

where $d(x_p, x_q)$ is the spatial distance between the vertices and $\sigma_c(x_p)$ is the standard deviation of the intensity in the cross section that $x_p$ located on.

Finally, we use the graph min-cut segmentation framework[25] to minimize the energy function (Eq. 5) and find the optimal surface separating the coronary artery lumen from its surrounding.

For each patient, we generated 3D models of the coronary tree with accounting for the PVE using Eq. 9, and without accounting for the PVE using a modified version of Eq. 8:

$$Pr(x_p \in lumen) = Pr_d(xp \in Lumen) \quad [12]$$

## 2.C. Flow simulation:

We estimated the hemodynamic significance of each lesion using the Lumped Parameter Model (LM) as proposed by Nickisch et al.[26] The LM represents the 3D coronary tree as a binary segment tree of vessel segments which in turn is translated into a nonlinear resistance network. Simple equations governing flow and friction in tubular structures are employed to simulate pressure drop and thus, non-invasive FFR estimates from CCTA. Both linear and non-linear resistors are used to represent the local pressure drop in coronary arteries and their bifurcations. As boundary conditions for flow simulation we employed an ostial pressure of $p = 100\ mmHg$ and outlet resistances $R_i$ scaling with the outlet diameter $d_i$[27]:

$$R_i \propto d_i^{-\frac{1}{3}}\ [13]$$

independently for each of the left and right coronary trees.

## 3. Evaluation

### 3.A. Methodology:

We implemented our main algorithm in C++ using the graph min-cut solver of Boykov et al[25], and an accelerated approximate K-nearest neighbor search.[24] We experimentally set the value of the regularization term $\lambda$ in Eq. 5 to 1.75 and $K$ in Eq. 6 to 100. The average running time to segment the entire coronary tree lumen for each patient was ~10 sec and the average running time for the flow simulation was less than 1 sec. for each coronary tree on a DELL T5550 Workstation equipped with 2 Intel® Xeon® x5650 at 2.66 GHz and 40GB RAM.

3.A.1. Comparison against the MICCAI 2012 challenge dataset

We first compare the performance of our segmentation algorithm with and without accounting for PVE to previously published approaches[11], including Lugauer et al[15,16], by means of segmentation accuracy using the 18 training datasets publicly available from the MICCAI 2012 challenge and evaluation framework[12]. A detailed description of the evaluation methodology is available in Kirişli et al. [11] We used the automatically generated centerlines provided by the method of Goldenberg et al. [21]

### 3.A.2. Parameter sensitivity analysis

Our segmentation algorithm includes several parameters that can be tuned. The most influential parameters that have effect on each cross-sectional contour are: $\lambda$ (Eq. 5) and $K$ (Eq. 6). We assessed the sensitivity of our algorithm to the two key parameters in our algorithm by using the training data available from the MICCAI 2012 challenge dataset. [12]

### 3.A.3. Impact of accounting for PVE on simulated FFR performance

Next, we assessed the performance of simulated FFR measurements based on automatically generated coronary 3D models in detecting significant CAD with invasive FFR measurement threshold of 0.8 as the reference standard by comparing the sensitivity, specificity, positive predictive value, negative predictive value, accuracy and overall area under the ROC curve for segmentations obtained using our algorithm with and without accounting for PVE.

We also evaluated specifically the potential benefit of accounting for PVE in automatic-segmentation for flow-simulation for obstructive lesions (CSA stenosis of 50% to 90%)[28] based on CCTA that are considered flow-limiting based on which patients are normally sent to invasive coronary angiography (ICA).

We determined the statistical significance of the improvement in the area under the ROC curve (AUC) achieved by accounting for the PVE using Delong's test.[29]

## 3.B Results:

### 3.B.1. Comparison against the MICCAI 2012 challenge dataset

Table 1 presents the segmentation accuracy results of our algorithm with and without accounting for PVE evaluated using the MICCAI 2012 challenge framework [11,12] in comparison with the results of Mohr et al[11] and Lugauer et al. [16] We refer the reader to the challenge website[12] for further comparison with the rest of the methods and with the observer performance.

**Table 1**: Summary statistics of coronary lumen segmentation accuracy using the MICCAI 2012 challenge evaluation framework[11,12] for the training datasets (18 cases, 78 coronary segments). Results presented for healthy and diseased segments separately and in the relevant metric units.

| Method | Category | Dice (%) | | MSD (mm) | | MAX SD (mm) | |
|---|---|---|---|---|---|---|---|
| | | Healthy | Disease | Healthy | Disease | Healthy | Disease |
| Lugauer et al.[16] | Automatic | 0.77 | 0.75 | 0.32 | 0.27 | 2.79 | 1.96 |
| Mohr et al[11] | Automatic | 0.75 | 0.73 | 0.45 | 0.29 | 3.73 | 1.87 |
| Automatic segmentation without accounting for PVE | Automatic | 0.69 | 0.74 | 0.5 | 0.28 | 1.67 | 1.3 |
| Automatic segmentation with accounting for PVE | Automatic | 0.69 | 0.74 | 0.49 | 0.28 | 1.69 | 1.22 |

### 3.B.2. Parameter sensitivity analysis

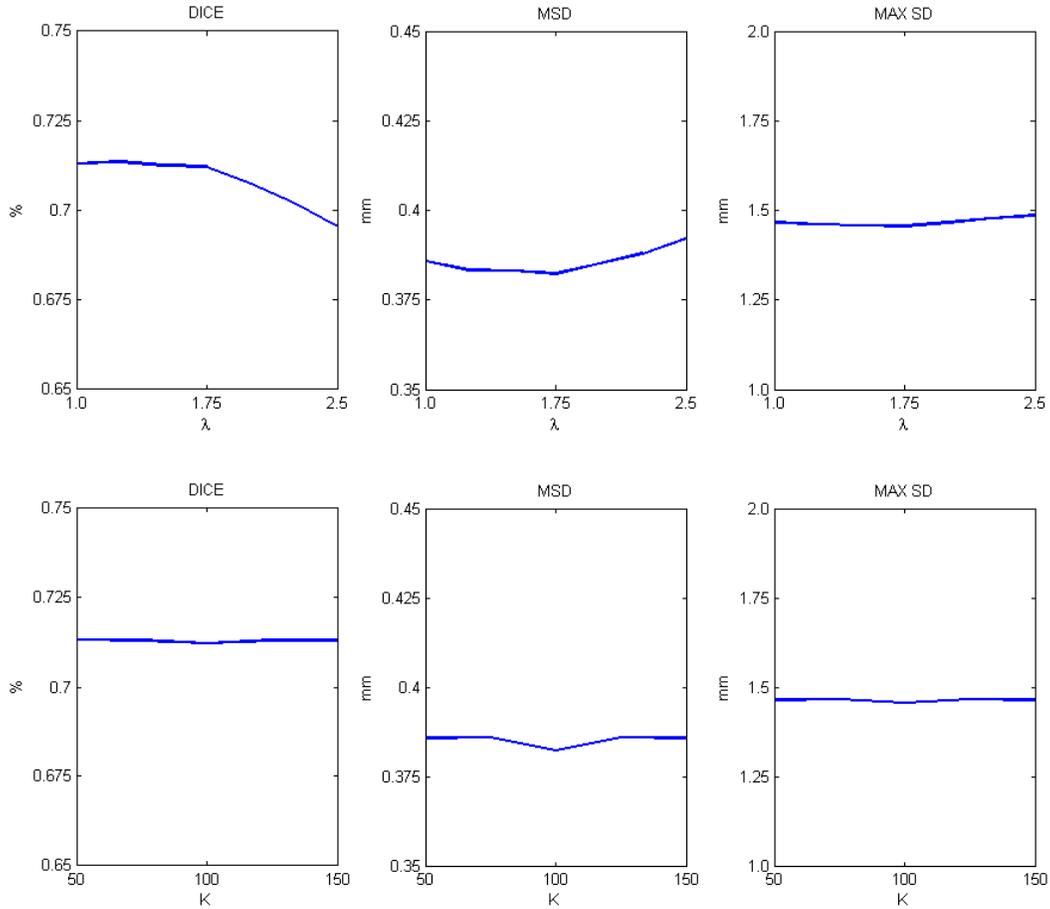

Fig. 4: Segmentation performance measures as a function of the two key parameters of the algorithm: $\lambda$ (Eq. 5, upper row) and $K$ (Eq. 6, bottom row), over the 18 training cases from the MICCAI 2012 dataset. [12] Measures used are the Dice coefficient (left column), the mean surface distance (mid column), and the maximal surface distance (right column).

Fig. 4 presents the variance in algorithm performance metrics as a function of the two key parameters of the algorithm: $\lambda$ (Eq. 5, upper row) and $K$ (Eq. 6, bottom row), over the 18 training cases from the MICCAI 2012 dataset. [12]   The algorithm is more sensitive to changes in $\lambda$ compared to changes in $K$. However, for both parameters, the differences in algorithm performance metrics as a function of the parameter values are small.

3.B.3. Impact of accounting for PVE on simulated FFR performance

Fig. 5 depicts representative examples of cross-sectional and straight multi-planar reconstructed images of coronary artery segmentation results with and without accounting for PVE along with the coronary centerline intensity profile used to detect the PVE.  Fig. 6 shows representative results of the 3D simulations in a color-coded mesh as obtained from the automatic coronary segmentation with and without accounting for the PVE along with the invasively measured FFR values.

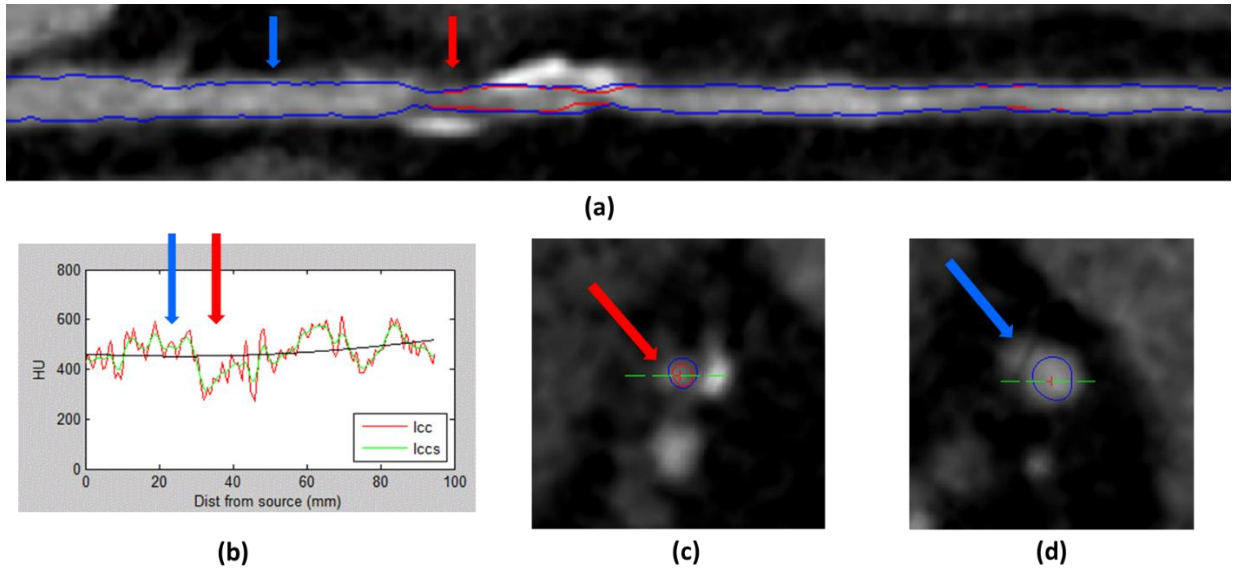

Fig. 5: Representative example of straight multi-planar reconstructed (a) and cross-sectional images (c-d) of coronary artery segmentation results with (red) and without (blue) accounting for the Partial Volume Effects (PVE) along with the coronary centerline intensity profile (b, red: original profile, green: smoothed version) used to detect the PVE. The blue arrow indicates location without PVE and the red arrow indicates a location with PVE.

Note the observed reduction in the HU due to the PVE.

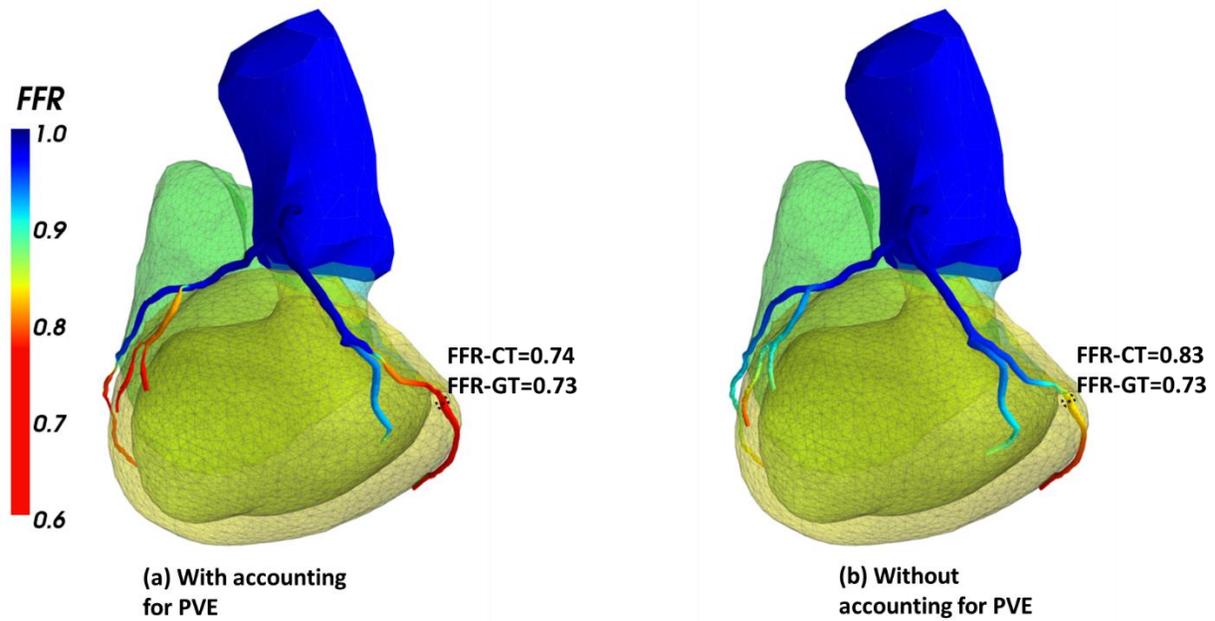

**Fig. 6:** Representative example of 3D color-coded visualization of the flow simulation results of the same case using coronary models generated with (a) and without (b) accounting for the Partial Volume Effects (PVE) along with measured and simulated FFR values after the lesion of interest. Accounting for PVE in the coronary segmentation phase yields more accurate FFR estimation.

Table 2 provides the summary statistics for the performance of CCTA based hemodynamic significance assessment using automatic coronary lumen segmentation with FFR≤0.8 as the reference for the entire set of coronary lesions. Integrating PVE analysis into automatic coronary lumen segmentation algorithm improved the specificity by 13.3% from 0.6 to 0.68 with the same sensitivity of 0.83. Also, accounting for PVE improved the area under the ROC curve for detecting hemodynamically significant CAD was improved from 0.76 to 0.8 compared to automatic segmentation without PVE analysis. The improvement in the AUC, however, did not reach the level of statistical significance (Delong's test[29], p=0.22).

**Table 2**: Summary statistics of flow simulation results with FFR≤0.8 as the reference for hemodynamic significance for entire dataset (N=132).

| | Sensitivity | Specificity | Accuracy | Negative predictive value | Positive predictive value | Area under the curve |
|---|---|---|---|---|---|---|
| Automatic segmentation with accounting for PVE | 0.83 | 0.68 | 0.74 | 0.88 | 0.6 | 0.8 |
| Automatic segmentation without accounting for PVE | 0.83 | 0.6 | 0.68 | 0.86 | 0.54 | 0.76 |

Table 3 presents the summary statistics for the performance of CCTA based hemodynamic significance assessment using automatic coronary lumen segmentation with FFR≤0.8 as the reference for lesions classified as obstructive (CSA stenosis 50% to 90%)[28] as diagnosed on CCTA. Accounting for PVE during the automatic coronary lumen segmentation algorithm improved specificity by ~43% from 0.51 to 0.73 with same sensitivity of 0.83 and the area under the curve by ~14% from 0.69 to 0.79. The improvement in the AUC was statistically significant (N=76, Delong's test[29], p=0.012).

**Table 3**: Summary statistics of flow simulation results with FFR≤0.8 as the reference for hemodynamic significance for obstructive lesions (CSA stenosis: 50%-90%, N=76).

|  | Sensitivity | Specificity | Accuracy | Negative predictive value | Positive predictive value | Area under the curve |
|---|---|---|---|---|---|---|
| Automatic segmentation with accounting for PVE | 0.83 | 0.73 | 0.78 | 0.83 | 0.73 | 0.79 |
| Automatic segmentation without accounting for PVE | 0.83 | 0.51 | 0.66 | 0.78 | 0.59 | 0.69 |

Fig. 7 presents the ROC curves for classifying coronary lesions as hemodynamically significant based on the flow simulation results using the automatically generated 3D models with and without accounting for PVE for the entire dataset (a) and for obstructive lesions (b).

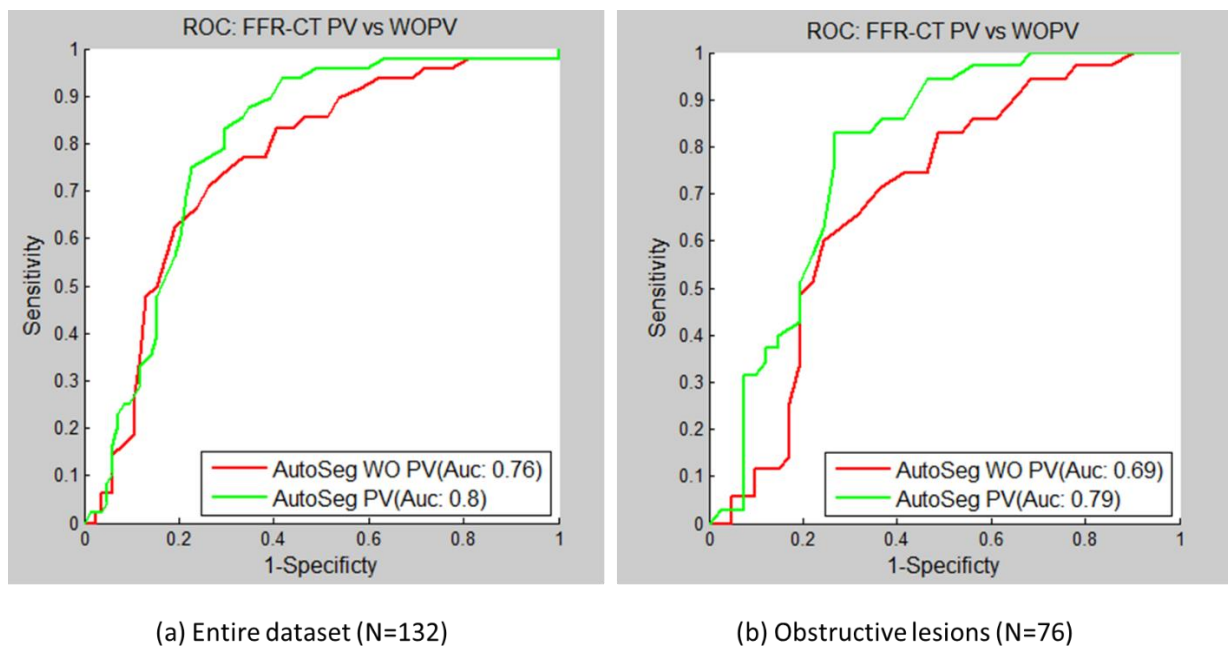

(a) Entire dataset (N=132)          (b) Obstructive lesions (N=76)

**Fig. 7:** ROC curves of cross-sectional area stenosis and flow simulation based on automatically generated 3D modes of the coronaries with and without accounting for the Partial Volume Effects (PVE) using invasive FFR≤0.8 as the reference for: (a) entire dataset, and (b) for lesions classified as obstructive (Cross Sectional Area (CSA) stenosis: 50%-90%).

## 4. <u>Discussion:</u>

Our study demonstrates the importance of accounting for PVE in automatic coronary segmentation algorithms used to determine the hemodynamic significance of coronary artery stenosis by CCTA based on flow simulations. Quantitative analysis of the CAD from CCTA required both automatic extraction of the coronaries' centerlines[30–34] and automatic segmentation of the coronaries lumen. Previous coronary lumen segmentation algorithm evaluation studies focused on the anatomical agreement between the automatic and manual segmentation. [11] The impact of accounting for PVE in automatic coronary lumen segmentation on functional assessment of coronary lesions, however, has not been previously explored.

In this work we have presented an algorithm for automatic coronary segmentation that accounts specifically for the PVE. Our algorithm detects locations potentially subjected to PVE by analyzing the intensity profile along the coronary centerline. Then, it incorporates this information into a machine-learning-based graph min-cut segmentation framework to obtain final 3D model of the coronary artery.

Our comparison with other previously published methods using the training data from the MICCAI 2012 evaluation framework show that our algorithm achieved the least maximal surface distance with a reduction of 38% for diseased segments and 40% for healthy segments. Both Dice and MSD measures are within the observer variability and comparable to previously published method of Mohr et al[11] using the same set of centerlines and slightly worse compared to Lugauer et al[16] which used a different set of centerlines as input . There was no substantial difference in performance of our method with and without accounting for PVE by means of segmentation accuracy. This may be attributed to the relatively small number of regions which may affected by PVE as detected by our algorithm compared to the overall number of evaluated segments.

In contrast, our flow simulation results demonstrate first that accounting for PVE can improve the accuracy of flow-simulation based assessment of coronary lesions' hemodynamic significance by 13.3%. In addition, the specific analysis of our method on obstructive lesions suggests that assessing the hemodynamic significance of coronary lesions using 3D models generated automatically using our algorithm has the potential to reduce the number of patients who otherwise will be scheduled for an invasive exam based on their CCTA results. The accuracy was similar compared to previous work wherein 3D models were corrected manually in a time consuming process. While the improvement by means of AUC was statistically significant for obstructive lesions it did not reach the significant level for entire dataset of 132 lesions. However, it is important to note that in the region of clinical interest (i.e. specificity>0.8) the method with PVE improved substantially over the method without PVE even for the entire dataset. We hypothesize that the reduction is due to some false-positive identification of PVE that reduces the overall specificity for high sensitivity FFR-CT thresholds.

The flow simulation results in the current study are in agreement with previous results presented at the SPIE 2016 Medical Imaging meeting by Freiman et al[20] in which accounting for potential PVE in automatic coronary lumen segmentation improves the performance of coronary lesions' hemodynamic significance assessment from CCTA data.

Several differences between the works should be noted. First, the work presented at the SPIE evaluated the impact of accounting for PVE using a limited number of datasets compared to the current work which includes additional cases with challenging lesions. The additional cases reduce the overall magnitude of the impact of accounting for PVE in the lumen segmentation. Second, the Matlab® prototype of the lumen segmentation algorithm used for the work presented at the SPIE replaced by an improved C++ implementation to achieve better performance. Finally, in our work presented at the SPIE, we performed the flow simulations using a finite-elements approach to solve the governing flow equations on a 3D coronary tree volumetric polyhedral mesh as demonstrated by Taylor et al.[35] In the current work we solved the governing flow equations using the lumped parameter model (LM) approach of Nicksich et al. [26] which translated the vessel tree into a network with simple equations governing flow and friction in tubular structures to determine pressure drop. The LM approach is computationally less intensive and more robust compared to the finite-elements approach while maintaining the same level of accuracy. [26]

Our study has several limitations: First, while the segmentation accuracy study was performed on a publicly available database with CCTA data from different vendors, the flow simulation study was limited to datasets obtained with CT scanners from one vendor. A more comprehensive study including CCTA data acquired by scanners from multiple vendors is desired to assess the full potential of automatic coronary tree segmentation in assessing the hemodynamic significance of coronary lesions. Second, our algorithm assumes a correct centerline of the coronary artery as input. Our segmentation accuracy experiment using the MICCAI 2012 database show that using automatically generated centerlines[21] as provided in this database is sufficient to obtain state-of-the art results using our method compared to other methods used the same centerlines[11]. However, the full impact of centerline extraction on the final goal of CCTA based hemodynamic significance assessment of CAD yet to be evaluated. Third, workflows for CCTA based hemodynamic significance assessment typically needs expert user interaction. Thus the added value of the coronary lumen segmentation that accounts for PVE should be further evaluated with expert corrections of the lumen contours where required to assess any further improvements in identifying lesions that are hemodynamically significant.

In conclusion, we have presented an automatic coronary lumen segmentation algorithm that accounts specifically for PVE. We have demonstrated the added value of accounting for PVE in assessing the hemodynamic significance of coronary lesions. The proposed algorithm has the potential to facilitate CCTA based hemodynamic significance assessment of CAD in clinical routine.

**Disclosure of Conflicts of Interest:**

M.F., M.V. and L.G. are employees of Philips Healthcare.

H.N., S.P. and H.S. are employees of Philips Research.

P.D has no conflict of interest related to this contribution.

P.M. has no conflict of interest related to this contribution.

**References:**

1.   Nowbar AN, Howard JP, Finegold JA, Asaria P, Francis DP. 2014 Global geographic analysis of mortality from ischaemic heart disease by country, age and income: Statistics from World Health Organisation and United Nations. *Int J Cardiol.* 2014;174(2):293-298. doi:10.1016/j.ijcard.2014.04.096.

2.   Scot-heart T. CT coronary angiography in patients with suspected angina due to coronary heart disease (SCOT-HEART): an open-label, parallel-group, multicentre trial. *Lancet.* 2015;385(9985):2383-2391. doi:10.1016/S0140-6736(15)60291-4.

3.   Meijboom WB, Van Mieghem CAG, van Pelt N, et al. Comprehensive Assessment of Coronary Artery Stenoses. Computed Tomography Coronary Angiography Versus Conventional Coronary Angiography and Correlation With Fractional Flow Reserve in Patients With Stable Angina. *J Am Coll Cardiol.* 2008;52(8):636-643. doi:10.1016/j.jacc.2008.05.024.

4.   Gonzalez JA, Lipinski MJ, Flors L, Shaw PW, Kramer CM, Salerno M. Meta-analysis of diagnostic performance of coronary computed tomography angiography, computed tomography perfusion, and computed tomography-fractional flow reserve in functional myocardial ischemia assessment versus invasive fractional flow reserve. *Am J Cardiol.* 2015;116(9):1469-1478. doi:10.1016/j.amjcard.2015.07.078.

5.   Nørgaard BL, Leipsic J, Gaur S, et al. Diagnostic performance of non-invasive fractional flow reserve derived from coronary CT angiography in suspected coronary artery disease: The NXT trial. *J Am Coll Cardiol.* 2014;63(12):1145-

1155. doi:10.1016/j.jacc.2013.11.043.

6.    Min JK, Taylor CA, Achenbach S, et al. Noninvasive fractional flow reserve derived from coronary CT angiography clinical data and scientific principles. *JACC Cardiovasc Imaging*. 2015;8(10):1209-1222. doi:10.1016/j.jcmg.2015.08.006.

7.    Min JK, Leipsic J, Pencina MJ, et al. Diagnostic accuracy of fractional flow reserve from anatomic CT angiography. *Jama*. 2012;308(12):1237-1245. doi:10.1001/2012.jama.11274.

8.    Coenen A, Lubbers MM, Kurata A, et al. Fractional flow reserve computed from noninvasive CT angiography data: diagnostic performance of an on-site clinician-operated computational fluid dynamics algorithm. *Radiology*. 2015;274(3):674-683. doi:10.1148/radiol.14140992.

9.    Kirbas, C. Quek F. A review of vessel extraction techniques and algorithms. *Comput Surv*. 2004;36(2):81-121. doi:10.1145/1031120.1031121.

10.   Lesage D, Angelini ED, Bloch I, Funka-Lea G. A review of 3D vessel lumen segmentation techniques: Models, features and extraction schemes. *Med Image Anal*. 2009;13(6):819-845. doi:10.1016/j.media.2009.07.011.

11.   Kirişli HA, Schaap M, Metz CT, et al. Standardized evaluation framework for evaluating coronary artery stenosis detection, stenosis quantification and lumen segmentation algorithms in computed tomography angiography. *Med Image Anal*. 2013;17(8):859-876. doi:10.1016/j.media.2013.05.007.

12.   Coronary Artery Stenoses Detection and Quantification Evaluation Framework. http://coronary.bigr.nl/stenoses/.

13.    Li KLK, Wu XWX, Chen DZ, Sonka M. Globally optimal segmentation of interacting surfaces with geometric constraints. *Proc 2004 IEEE Comput Soc Conf Comput Vis Pattern Recognition, 2004 CVPR 2004*. 2004;1:0-5. doi:10.1109/CVPR.2004.1315059.

14.    Veni G, Fu Z, Awate SP, Whitaker RT. Bayesian Segmentation of Atrium Wall Using Globally-Optimal Graph Cuts on 3D Meshes. In: Gee JC, Sarang J, Kilian MP, Wells WM, Zöllei L, EDS. *Inf Process Med Imaging. LNCS, Vol.* 7917. Springer-Verlag Berlin, Heidelberg; 2013:656–667. doi: 10.1007/978-3-642-38868-2_55

15.    Lugauer F, Zhang J, Zheng Y, Hornegger J, Kelm BM. Improving accuracy in coronary lumen segmentation via explicit calcium exclusion, learning-based ray detection and surface optimization. *SPIE Med Imaging*. 2014;9034:90343U-90343U-10. doi:10.1117/12.2043238.

16.    Lugauer F, Zheng Y, Hornegger J, Kelm BM. Precise lumen segmentation in coronary computed tomography angiography. In: Menze B, Langs G, Montillo A, et al, eds. *Medical Computer Vision: Algorithms for Big Data LNCS, Vol.* 8848. Springer-Verlag Berlin, Heidelberg; 2014:137–147. doi:10.1007/978-3-319-13972-2_13.

17.    Sato Y, Yamamoto S, Tamura S. Accurate Quantification of Small-Diameter Tubular Structures in Isotropic CT Volume Data Based on Multiscale Line Filter Responses. In: Barillot C, Haynor DR, and Hellier P, eds. *Medical Image Computing and Computer-Assisted Intervention -- MICCAI 2004: 7th International Conference, LNCS, Vol.* 3216. Springer International Publishing; 2004:508-515.

doi: 10.1007/978-3-540-30135-6_62.

18.  Prevrhal S, Fox JC, Shepherd JA, Genant HK. Accuracy of CT-based thickness measurement of thin structures: modeling of limited spatial resolution in all three dimensions. *Med Phys*. 2003;30(1):1-8. doi:10.1118/1.1521940.

19.  Varma JK, Subramanyan K, Durgan J. Full width at half maximum as a measure of vessel diameter in computed tomography angiography. In: Vol 5372. ; 2004:447-454. http://dx.doi.org/10.1117/12.535642.

20.  Freiman M, Lamash Y, Gilboa G, et al. Automatic coronary lumen segmentation with partial volume modeling improves lesions' hemodynamic significance assessment. In: *SPIE Medical Imaging*. ; 2016:978403. http://dx.doi.org/10.1117/12.2209476

21.  Goldenberg R, Eilot D, Begelman G, Walach E, Ben-Ishai E, Peled N. Computer-aided simple triage (CAST) for coronary CT angiography (CCTA). *Int J Comput Assist Radiol Surg*. 2012;7(6):819-827. doi:10.1007/s11548-012-0684-7.

22.  Shahzad R, Kirişli H, Metz C, et al. Automatic segmentation, detection and quantification of coronary artery stenoses on CTA. *Int J Cardiovasc Imaging*. 2013:1847-1859. doi:10.1007/s10554-013-0271-1.

23.  Cover T, Hart P. Nearest neighbor pattern classification. *IEEE Trans Inf Theory*. 1967;13(1):21-27. doi:10.1109/TIT.1967.1053964.

24.  Muja M, Lowe DG. Fast Approximate Nearest Neighbors with Automatic Algorithm Configuration. *Int Conf Comput Vis Theory Appl (VISAPP '09)*. 2009:1-10. doi:10.1.1.160.1721.

25.  Boykov Y, Funka-Lea G. Graph cuts and efficient N-D image segmentation. *Int J*

*Comput Vis.* 2006;70(2):109-131. doi:10.1007/s11263-006-7934-5.

26.    Nickisch H, Lamash Y, Prevrhal S, et al. Learning Patient-Specific Lumped

       Models for Interactive Coronary Blood Flow Simulations. In: Navab N, Hornegger

       J, Wells MW, Frangi FA, eds. *Medical Image Computing and Computer-Assisted*

       *Intervention -- MICCAI 2015: 18th International Conference, LNCS, Vol. 9350*.

       Springer-Verlag Berlin, Heidelberg; 2015:433-441. doi:10.1007/978-3-319-24571-

       3_52.

27.    Huo Y, Kassab GS. Intraspecific scaling laws of vascular trees. *J R Soc Interface*.

       2012;9(66):190-200. doi:10.1098/rsif.2011.0270.

28.    Kruk M, Wardziak Ł, Demkow M, et al. Workstation-Based Calculation of CTA-

       Based FFR for Intermediate Stenosis. *JACC Cardiovasc Imaging*. 2016;9(6):690-

       699. doi:10.1016/j.jcmg.2015.09.019.

29.    Delong ER, DeLong DM, Clarke-Pearson DL. Comparing the areas under two or

       more correlated receiver operating characteristic curves: a nonparametric

       approach. *Biometrics*. 1988;44(3):837-845. doi:10.2307/2531595.

30.    Ecabert O, Peters J, Walker MJ, et al. Segmentation of the heart and great

       vessels in CT images using a model-based adaptation framework. *Med Image*

       *Anal*. 2011;15(6):863-876. doi:10.1016/j.media.2011.06.004.

31.    Ecabert O, Peters J, Schramm H, et al. Automatic model-based segmentation of

       the heart in CT images. *IEEE Trans Med Imaging*. 2008;27(9):1189-1202.

       doi:10.1109/TMI.2008.918330.

32.    Freiman M, Joskowicz L, Broide N, et al. Carotid vasculature modeling from

       patient CT angiography studies for interventional procedures simulation. *Int J*

*Comput Assist Radiol Surg.* 2012;7(5):799-812. doi:10.1007/s11548-012-0673-x.

33. Zheng Y, Tek H, Funka-Lea G. Robust and accurate coronary artery centerline extraction in CTA by combining model-driven and data-driven approaches. In: Mori K, Sakuma I, Sato Y, Barillot C, and Navab N, eds. *Medical Image Computing and Computer-Assisted Intervention -- MICCAI 2013: 16th International Conference, LNCS, Vol. 8151.* Springer-Verlag Berlin, Heidelberg; 2013:74-81. doi:10.1007/978-3-642-40760-4_10.

34. Schaap M, Metz CT, van Walsum T, et al. Standardized evaluation methodology and reference database for evaluating coronary artery centerline extraction algorithms. *Med Image Anal.* 2009;13(5):701-714. doi:10.1016/j.media.2009.06.003.

35. Taylor CA, Fonte TA, Min JK. Computational fluid dynamics applied to cardiac computed tomography for noninvasive quantification of fractional flow reserve: Scientific basis. *J Am Coll Cardiol.* 2013;61(22):2233-2241. doi:10.1016/j.jacc.2012.11.083.